\documentclass{pasj00}

\begin{document}
\SetRunningHead{Niino et al.}{GRB~100418A}

\title{GRB~100418A: a Long GRB without a Bright Supernova 
in a High-Metallicity Host Galaxy\thanks{Based on data collected at Subaru Telescope, 
which is operated by the National Astronomical Observatory of Japan.}}

\author{Yuu \textsc{Niino}, Tetsuya \textsc{Hashimoto}}
\affil{Division of Optical and IR Astronomy, National Astronomical Observatory of Japan, 
2-21-1 Osawa, Mitaka, Tokyo}
\email{yuu.niino@nao.ac.jp}

\author{Kentaro \textsc{Aoki}, Takashi \textsc{Hattori}} 
\affil{Subaru Telescope, National Astronomical Observatory of Japan\\
650 North A`ohoku Place, Hilo, HI 96720, U.S.A.}

\author{Kiyoto \textsc{Yabe}\thanks
{Present address: Division of Optical and IR Astronomy, National Astronomical Observatory of Japan, 2-21-1 Osawa, Mitaka, Tokyo}}
\affil{Department of Astronomy, Kyoto University, Kitashirakawa-Oiwakecho, Sakyo-ku, Kyoto}

\and

\author{Ken'ichi {\sc Nomoto}}
\affil{Institute for the Physics and Mathematics of the Universe, The University of Tokyo\\ 
5-1-5 Kashiwanoha, Kashiwa, Chiba}


\maketitle

\begin{abstract}
We present results of a search for a supernova (SN) component 
associated with GRB~100418A at the redshift of 0.624. 
The field of GRB~100418A was observed with FOCAS on Subaru 8.2m telescope 
under a photometric condition (seeing \timeform{0.3''}--\timeform{0.4''}) on 2010 May 14 (UT). 
The date corresponds to 25.6 days after the burst trigger (15.8 days in the restframe). 
We did imaging observations in $V,\ R_{c}$, and $I_{c}$ bands, 
and two hours of spectrophotometric observations. 
We got the resolved host galaxy image which elongated \timeform{1.6''} (= 11 kpc) from north to south. 
No point source was detected on the host galaxy. 
The time variation of $R_{c}-$band magnitude shows 
that the afterglow of GRB~100418A has faded 
to $R_c \gtrsim 24$ without SN like rebrightening, 
when we compare our measurement to the reports in GCN circulars. 
We could not identify any SN feature 
such as broad emission-lines or bumps in our spectrum. 
Assuming the SN is fainter than the $3\sigma$ noise spectrum of our observation, 
we estimate the upper limit on the SN absolute magnitude $M_{Ic,{\rm obs}}>-17.2$ in observer frame $I_c-$band. 
This magnitude is comparable to the faintest type Ic SNe. 
We also estimate host galaxy properties from the spectrum. 
The host galaxy of GRB~100418A is relatively massive (log~$M_\star/M_\odot = 9.54$) 
compared to typical long GRB host galaxies, and has 12+log(O/H) = 8.75. 
\end{abstract}

\section{Introduction}
\label{sec:Intro}

Long gamma-ray bursts (GRBs) are now considered to be death of massive stars (so called collapsar scenario). 
The most convincing observational evidences have been associations with supernovae (SNe).
Some associations were spectroscopically, others were photometrically. 
However, there is at least one long GRB, GRB~060614 
(duration $\sim 100$ sec, \cite{DellaValle:06a,Fynbo:06a})
whose SN component was not detectable to very deep limits. 
GRB~060505 is also a $T_{90} > 2$ sec burst without a detectable SN \citep{Fynbo:06a}, 
although its duration $T_{90} =$ 4 sec is close to the classical 2 sec threshold 
which separates the long and short populations of GRBs. 

Various progenitor models have been proposed to explain these events. 
Some suggested they are short GRBs, 
which are considered to originate in mergers of double compact object binary, 
with longer duration than typical short GRBs \citep{Gehrels:06a,Ofek:07a,Levesque:07a,Caito:09a}, 
while others point out similarities of these GRBs to other long bursts 
which originate from collapsars \citep{McBreen:08a,Thone:08a,Xu:09a}. 
The possibility of new populations which are different from long and short GRBs 
are also suggested for these GRBs \citep{Gal-Yam:06a,Lu:08a}. 
The sample of long GRBs with strong constraints on their SN components is still small, 
and it is still uncertain what impact do the long GRBs without SNe have 
upon our understanding of the GRB populations. 

Here we present results of a search for a SN component associated with GRB~100418A. 
GRB~100418A was triggered at 21:10:08 UT April 2010 \citep{Marshall:10a}. 
The duration and the energetics of the prompt emission were $T_{90}=8 \pm 2$ second 
and $E_{iso}=9.9^{+6.3}_{-3.4} \times 10^{50}$ erg \citep{Marshall:11a}. 
Follow-up spectroscopies of the optical transient (OT) detected emission-lines and absorption lines
with the common redshift of 0.624 \citep{Antonelli:10a, Cucchiara:10a}. 
\citet{Holland:10a} pointed out that the light curve of GRB~100418A is similar to previous GRBs with SNe. 
\citet{Marshall:10b} reported flattening of the optical light curve after 700 ks. 

Throughout this paper, we assume the fiducial cosmology with the Hubble constant 
of 71 km s$^{-1}$ Mpc$^{-1}$, $\Omega_{\Lambda}=0.73$ and $\Omega_{m}=0.27$.
The angular scale is thus 6.8 kpc/arcsec. 
We use the AB magnitude system unless otherwise stated. 

\section{Observations and Reductions}
\label{sec:Obs}

We observed the field of GRB~100418A with FOCAS \citep{Kashikawa:02a} on Subaru Telescope \citep{Iye:04a}
under a photometric condition (seeing \timeform{0.3''}--\timeform{0.4''}) on 2010 May 14 (UT).
We started imaging observations in $V,\ R_c$, and $I_c$ bands at 12:13 UT, and 
spectroscopic observations followed.
The log of the observations is tabulated in table~\ref{tb:observation}. 
We acquired a single exposure of 120 seconds in each of $V,\ R_c$, and $I_c$ bands 
with 2$\times$2 binning which yields pixel scale of \timeform{0.206''}/pixel.
Unbinned $R_c-$band images were also taken with \timeform{0.103''}/pixel.

Spectroscopy was performed with \timeform{0.8''} width slit 
and 3 pixels binned along the spatial direction, 
and two different settings of grisms and order-cut filter.
One setting covered a wavelength range between  
5800~{\AA} and 10200~{\AA} with the R300 grism and the O58 order-cut filter, 
and the other covered between 3900~{\AA} and 8300~{\AA} with the B300 grism and no order-cut filter.
The total integration time was 1 hour (1200 sec x 3) for each of the setting.
We used the atmospheric dispersion corrector, 
and set the slit position angle to {$0.0\deg$}, i.e., north-south.
The spectral resolution is $\sim$ 11~\AA\ in both settings.

The data were reduced using IRAF\footnote {IRAF is distributed by the
  National Optical Astronomy Observatory, which is operated by the
  Association of Universities for Research in Astronomy (AURA), Inc.,
  under cooperative agreement with the National Science Foundation.  }
for the procedures of bias subtraction and flat-fielding. 
The spectra were wavelength calibrated, and sky subtracted.  
Wavelength calibration was performed 
using night sky emission lines for the red setting and ThAr arc lines for the blue setting. 
and the rms wavelength calibration error is 0.2--0.3~\AA. 
The sensitivity calibration was performed as a function of wavelength 
by using the spectrum of Feige 34 observed with \timeform{2''} width slit. 
For imaging data, the flux calibration was carried out 
by observing the standard star PG~1505-027 field \citep{Stetson:00a} at the same night.

\begin{figure}
  \begin{center}
    \FigureFile(80mm,80mm){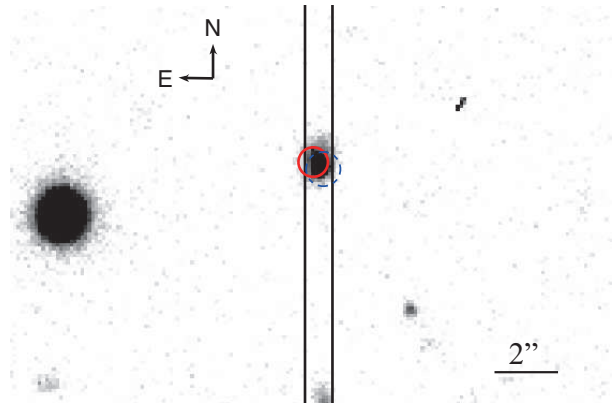}
  \end{center}
  \caption{The $R_c-$band image of the GRB~100418A host galaxy (unbinned, \timeform{0.103''}/pixel). 
  The circles indicate localizations of the optical counterpart in earlier epoch observations 
  by UVOT \citep[red solid]{Marshall:11a} and GROND \citep[blue dashed]{Filgas:10a}, respectively. 
  The slit position for our spectroscopy is shown with vertical lines. 
  A colored version of the figure is available in the online journal. }\label{fig:R1x1}
\end{figure}

\begin{figure*}
  \begin{center}
    \FigureFile(160mm,80mm){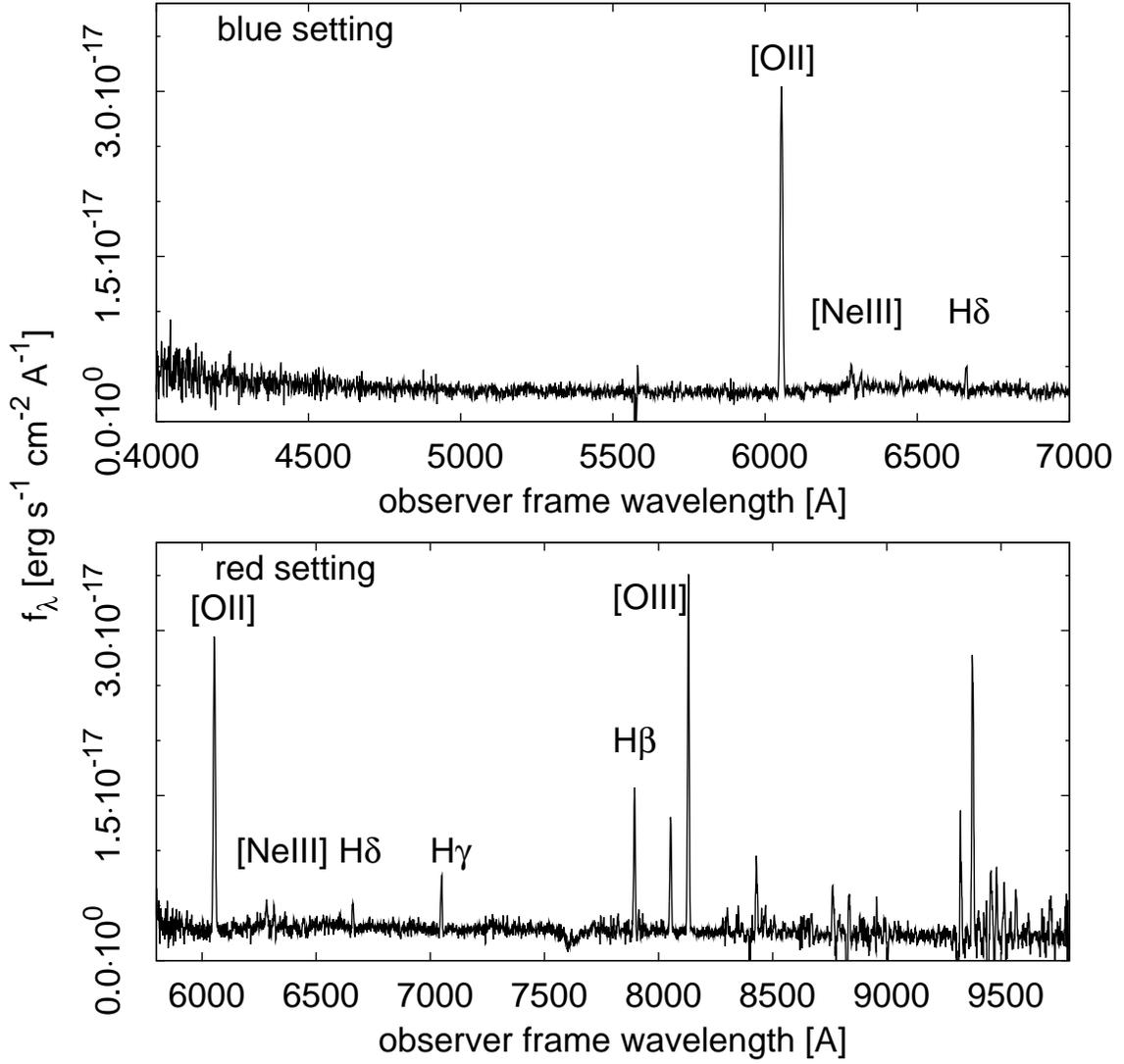}
  \end{center}
  \caption{The observed spectrum of the GRB~100418A host galaxy.}\label{fig:fullspectra}
\end{figure*}

The $R_{c}-$band image of \timeform{0.103''}/pixel is shown in figure~\ref{fig:R1x1}, 
and the observed spectra are shown in figure~\ref{fig:fullspectra}. 
These spectra are made by summing \timeform{2.5''} along the slit in order to include whole galaxy light.
We found strong emission lines such as [O\emissiontype{II}], H$\gamma$, H$\beta$, [O\emissiontype{III}]  
which are redshifted 0.624 consistently to the previous reports \citep{Antonelli:10a, Cucchiara:10a}.

\begin{table*}
 \caption{Observation logs}\label{tb:observation}
 \begin{center}
   \begin{tabular}{cccc}
     \hline
     UT date & time after the burst & filter & exposure time \\ 
midpoint of exposures & (days)& & (sec) \\
     \hline
     14.50969 & 25.62765 & $Rc$ & 120 \\
     14.51178 & 25.62974 & $Ic$ & 120 \\
     14.51388 & 25.63184 & $V$  & 120 \\
     14.51705 & 25.63501 & (Unbinned) $Rc$ & 120 $\times$ 2 \\
     14.56639 & 25.68435 & 300R+O58 & 1200 $\times$ 3 \\
     14.57130 & 25.68926 & 300B & 1200 $\times$ 3 \\
     \hline
   \end{tabular}
 \end{center}
\end{table*}

\section{Constraints on the Associated Supernova}
\subsection{Time Variation of Broad-Band Magnitudes}
\label{sec:timevariation}

We detected the host galaxy which has been catalogued in Sloan Digital Sky Survey (SDSS) 
as $g=22.89\pm0.17$, $r=22.41 \pm 0.16$, $i=21.94 \pm 0.17$ \citep{Malesani:10a}. 
The host galaxy is resolved in our images. 
It elongates \timeform{1.6''} (= 11 kpc) from north to south.
A bright spot locates at the southern tip of the galaxy.
The size of this bright spot is \timeform{0.47''} FWHM while stars' FWHMs are \timeform{0.32''}.
The spot is clearly extended and not a point source.
The aperture photometry of the host galaxy 
gives $V = 22.61\pm0.06$, $R_c = 22.10\pm0.03$, and $I_c = 21.80\pm0.05$. 
Our $R_c$-band magnitude is consistent to that 
reported by \citet{Rumyantsev:10a}, $R=22.06 \pm 0.06$ at 26.0 days after the burst.

\begin{figure}
  \begin{center}
    \FigureFile(80mm,80mm){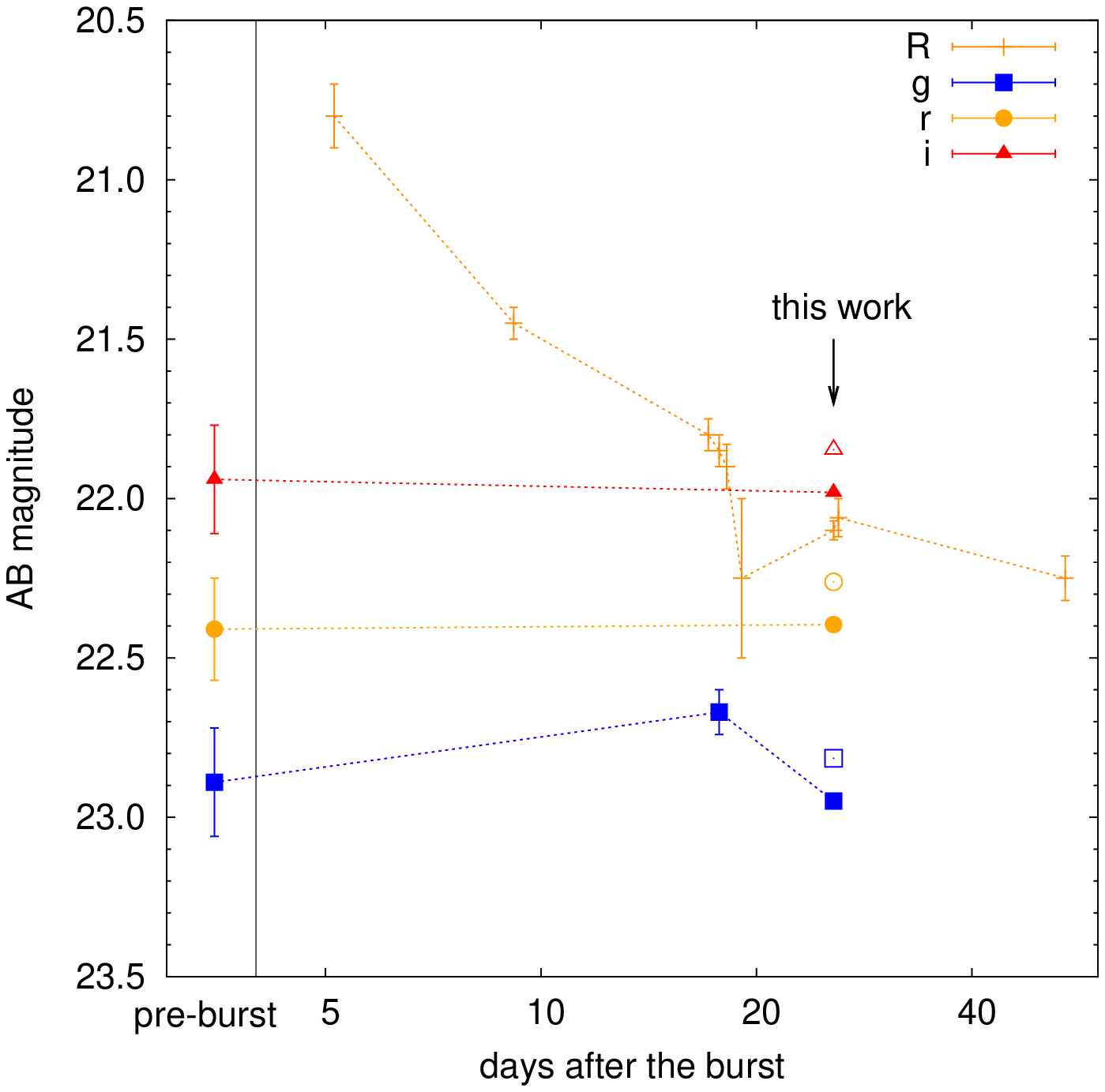}
  \end{center}
  \caption{The time variation of broad-band photometry 
  in GRB~100418A optical follow up observations (afterglow, host galaxy, and potential supernova). 
  The open symbols for $g,\ r$, and $i-$band indicates the correction for the possible slit loss. 
  We collect data points other than our observation from the GCN circulars 
  \citep{Bikmaev:10a,Bikmaev:10b,Malesani:10a,Perley:10a,Rumyantsev:10a,Rumyantsev:10b,Volnova:10a}. 
  A colored version of the figure is available in the online journal. }\label{fig:lightcurve}
\end{figure}

Time variation of $R-$band magnitude between 5 and 54 days after the burst is shown in figure~\ref{fig:lightcurve}, 
together with the $g,\ r$, and $i-$band magnitudes 
in the SDSS catalog (pre-burst) and those at 25.6 days after the burst. 
The $g,\ r,$ and $i-$band magnitudes at 25.6 d 
($g = 22.95, r = 22.40,$ and $i = 21.98$ with photometric errors $< 0.01$ mag) 
are calculated from our spectrum of the host galaxy following the equation (3) in \citet{Smith:02a}. 
The $R-$band time variation shows smooth decline of afterglow 
expect one datapoint with large error bar at 19.1 days after the burst. 
SN like rebrightening is not recognizable. 

The $g,\ r$, and $i-$band magnitudes calculated from the FOCAS spectrum 
are consistent to SDSS pre-burst magnitudes. 
However, it should be noted that $R_c-$band magnitude is 0.13 mag fainter 
when calculated from the spectra than that in the aperture photometry, 
possibly due to loss of light at the slit. 
The spectrum is still consistent to the pre-burst magnitudes 
within SDSS photometric error when 0.13 mag slit loss correction is applied, 
however it is systematically brighter in all of the three bands. 
\citet{Rumyantsev:10b} reported $R=22.25\pm0.07$ at 54.0 days after the burst, 
which is 0.15 mag fainter than our photometry at 25.6 d. 
Thus it is possible our image and spectrum contains 
small fraction of afterglow (and potential supernova) light, 
although no point source is detected on the host galaxy as described above. 
We compare the calculated $i-$band magnitude with the SDSS photometry after the slit loss correction, 
and obtain upper limit on the OT, $M_{i,{\rm obs}} > -19.1$ (95\% limit, afterglow plus potential supernova). 
Note that observer frame $i-$band covers $\sim 4200$--5300~\AA\ in the restframe. 

\subsection{Bright Spot Spectrum}

\begin{figure}
  \begin{center}
    \FigureFile(80mm,80mm){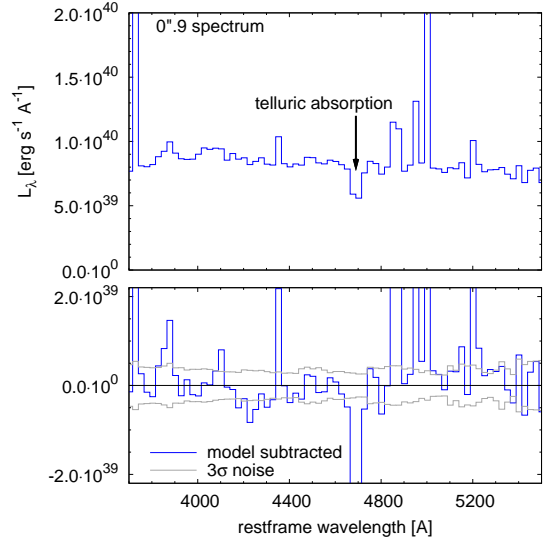}
  \end{center}
  \caption{{\it Top panel}: the observed spectra at the brightest \timeform{0.9''} of the host galaxy. 
  {\it Bottom panel}: the model subtracted \timeform{0.9''} spectrum (thick histogram) 
  compared to $3\sigma$ noise spectrum (thin histogram). }\label{fig:nucspectra}
\end{figure}

The OT position reported by \citet{Filgas:10a} and \citet{Marshall:11a} 
corresponds to the brightest spot of the host galaxy (figure~\ref{fig:R1x1}). 
To search SN features in our spectra, 
we extract spectrum of the brightest part (\timeform{0.9''}) of the host galaxy 
which corresponds to the error circles of the OT localization. 
The \timeform{0.9''} spectrum is shown in the top panel of figure~\ref{fig:nucspectra}. 
The spectrum is 30 pixels binned along the wavelength ($= 40.2$--40.5~\AA\ in observerframe), 
in order to increase signal to noise ratio (S/N). 
The \timeform{0.9''} spectrum shows no SN feature like broad emission-lines or bumps. 
We subtract a stellar spectral energy distribution (SED) model 
of the host galaxy (see \S\ref{sec:galprop}) from the \timeform{0.9''} spectrum.  
Because the SED model is fitted to the whole galaxy spectrum, 
it is scaled 75\% to match the \timeform{0.9''} spectrum. 
The subtracted spectrum is shown in the bottom panel of figure~\ref{fig:nucspectra}, 
together with $3\sigma$ noise spectrum. 
Although it is possible that the afterglow and the SN affect the SED model, 
and our spectra suffers from residulals of OH sky lines 
showing larger fluctuation than expected from the noise level,  
it shows no broad feature ($\gtrsim 200$~\AA) above the $3\sigma$ noise. 
Thus we estimate upper limit of the SN light assuming it is fainter than the $3\sigma$ noise spectrum. 

\begin{figure}
  \begin{center}
    \FigureFile(80mm,80mm){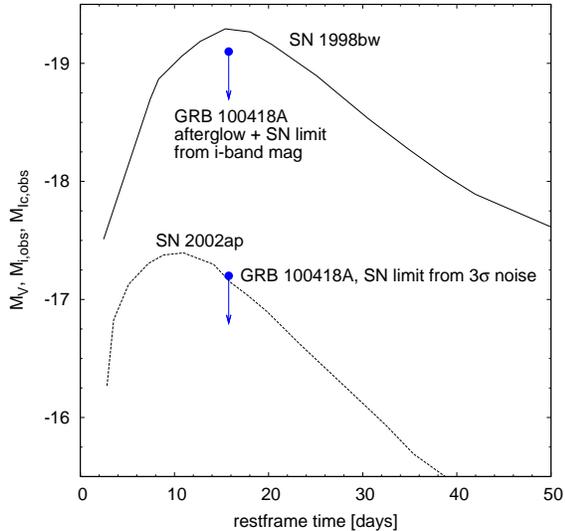}
  \end{center}
  \caption{Constraint on SN component associated with GRB~100418A. 
  Observer frame $i-$band limit from broad-band photometry (afterglow + SN), 
  and observer frame $I_c-$band limit from the $3\sigma$ noise spectrum (SN only) 
  is compared to $V-$band light curves of broad lined type Ic SNe in literature
  \citep{Galama:98a,Mazzali:02a}. }\label{fig:SNlimit}
\end{figure}

Spectrum of a SN ($L_{\lambda}$) peaks at 5000--5500~\AA\ in its restframe, 
and hence $V-$band magnitude is popularly used to discuss SN brightness at lower redshifts. 
However our spectrum suffers from strong OH sky lines in higher-end of the restframe $V-$band 
($\gtrsim 9000$~\AA\ in observer frame, see figure~\ref{fig:fullspectra}), and S/N is poor. 
Instead, we calculate constraint on the SN component 
in observer frame $I_c-$band which also covers restframe 5000--5500~\AA, 
and consider it is comparable to $V-$band magnitudes of other SNe at lower redshfits. 
We note that the line-of-sight extinction of GRB~100418A within its host galaxy is small, $E(B-V) = 0.056$, 
based on X-ray afterglow spectrum \citep{Marshall:11a}. 
The $3\sigma$ noise spectrum is equivalent to 
absolute magnitude $M_{Ic,{\rm obs}}=-17.2$ after corrected for the extinction (figure~\ref{fig:SNlimit}). 

\section{Properties of the Host Galaxy}
\label{sec:galprop}

\begin{figure}
  \begin{center}
    \FigureFile(80mm,80mm){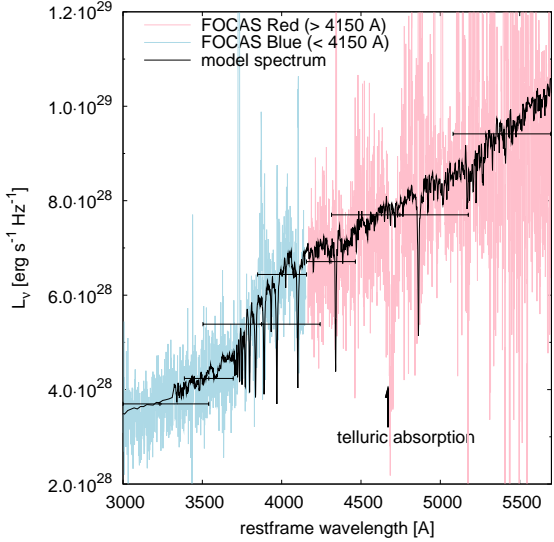}
  \end{center}
  \caption{The best fit SED model (thick black line) 
  is compared to observed spectrum (thin light-colored line), 
  and corresponding band magnitudes to which the fitting is performed. 
  The observed spectra is taken from the blue setting spectroscopy below 4150~\AA, 
  and from the red setting above (light-blue and pink in colored version). 
  A colored version of the figure is available in the online journal. }\label{fig:SEDfit}
\end{figure}

We perform SED fitting using {\it SEDfit} software package \citep{Sawicki:12a,Yabe:09a}, 
which utilize population synthesis models \citet{Bruzual:03a} and the extinction law by \citet{Calzetti:00a}. 
After corrected for the Milky Way extinction $E(B-V) = 0.072$ \citep{Schlegel:98a}, 
the spectrum is converted to magnitudes in 8 bands shown in figure~\ref{fig:SEDfit}, 
collected from both of the blue and red settings of the spectroscopy 
avoiding wavelength range where detector sensitivity is poor and/or OH sky line is strong. 
Stellar metallicity is assumed to be $Z_\odot$ for consistency 
with the emission line diagnostic discussed below. 
However, we note that the results with $0.2 Z_\odot$ model is not significantly different. 
We examine seven cases of star formation history: 
simple stellar population (SSP), constant star formation, 
and exponentially decaying star formation with $\tau = 10^{-3},\ 10^{-2},\ 0.1,\ 1,$ and 10 Gyr, 
among which SSP provides best fit. 

We find that the SED model reproduces our spectrum with parameters listed in table~\ref{tb:SEDparameter} 
[the initial mass function (IMF) of \citet{Salpeter:55a} is assumed]. 
The best fitting SED model is shown in figure~\ref{fig:SEDfit}. 
As discussed in \S\ref{sec:timevariation}, our spectrum may be contaminated by GRB afterglow. 
We also try SED fitting subtracting afterglow model from our spectrum, 
however the results are similar to the case without afterglow subtraction. 
The assumed afterglow model is 
that the late time spectral index of GRB~100418A afterglow $\beta=1.15$ \citep{Marshall:11a}, 
and the flux in $R-$band is $F_\nu = 6.8\times10^{-30}$ erg s$^{-1}$ cm$^2$ Hz$^{-1}$ 
which is derived from the difference between the $R-$band magnitudes at 25.6 d and 54.0 d (see figure~\ref{fig:lightcurve}). 
The results of the afterglow subtracted SED fitting is included in the errors shown in table~\ref{tb:SEDparameter}. 

\begin{table}
  \caption{The results of SED fitting.}\label{tb:SEDparameter}
  \begin{center}
    \begin{tabular}{cccc}
      \hline
      log~$M_\star$ & age [Gyr] & $E(B-V)$ & $\chi^2$ \\ 
      \hline
      $9.54^{+0.28}_{-0.03}$ & 7.1--8.1 & $0.38^{+0.06}_{-0.09}$ & 13.1 \\ 
      \hline
    \end{tabular}
  \end{center}
\end{table}

We measure metallicity of the host galaxy using emission line diagnostics. 
We perform linear continuum plus single gaussian fit to each line,
after subtracting the best fit SED model (figure~\ref{fig:SEDfit}) from observed spectrum, 
to minimize the effect of stellar absorption features such as Balmer absorption lines.
The best fit SED is obtained assuming stellar metallicity $Z_\odot$, 
but using SED model with $0.2 Z_\odot$ does not make significant difference. 
The resulting line fluxes are listed in table~\ref{tb:lines}. 
The H$\beta$/H$\gamma$ ratio is consistent to zero extinction, 
thus we assume there is no reddening on emission lines except that in the Milky Way. 

\begin{table*}
  \caption{The emission line fluxes of GRB~100418A host galaxy [10$^{-17}$ erg s$^{-1}$ cm$^{-2}$].}\label{tb:lines}
  \begin{center}
    \begin{tabular}{cccccccc}
      \hline
      [O\emissiontype{II}]3727 & [Ne\emissiontype{III}]3869 & H$\gamma$ & H$\beta$ & [O\emissiontype{III}]4959 & [O\emissiontype{III}]5007 & log~$O_{32}$ & log~$R_{23}$ \\ 
      \hline
      25.1$\pm0.2$ & 1.49$\pm0.16$ & 4.30$\pm0.10$ & 
      9.10$\pm0.18$ & 7.48$\pm0.12$ & 22.4$\pm0.2$ & 0.0757 & 0.781 \\ 
      \hline
    \end{tabular}
  \end{center}
\end{table*}

We use the $R_{23}$ method described in \citet{Kobulnicky:04a}, 
which gives two solutions of possible metallicity (upper and lower branch). 
\begin{eqnarray*}
12+{\rm log(O/H)}_{\rm upper} = 8.75 \\
12+{\rm log(O/H)}_{\rm lower} = 7.94
\end{eqnarray*}
\citet{Nagao:06a} found [Ne\emissiontype{III}]3869/[O\emissiontype{II}]3727 line flux ratio correlates with metallicity, 
and thus can be used as a indicator to separate these two solutions. 
They showed that galaxies with 12+log(O/H) $\geq$ 8.65 have log~[Ne\emissiontype{III}]3869/[O\emissiontype{II}]3727 $< -1.1$, 
while galaxies with 12+log(O/H) $< 8.05$ have log~[Ne\emissiontype{III}]3869/[O\emissiontype{II}]3727 $> -0.79$. 
The host galaxy of GRB~100418A has log~[Ne\emissiontype{III}]3869/[O\emissiontype{II}]3727 = $-1.23$, 
which suggests that the upper branch solution is the case. 

There are several emission line diagnostics proposed to measure metallicity, 
and their results are not always consistent to each other (e.g. \cite{Kewley:08a}). 
We note that metallicities discussed in \citet{Nagao:06a} is measuered 
using the \citet{Tremonti:04a} method for high-metallicity galaxies, 
and the electron temperature method \citep{Izotov:06a} for low-metallicity galaxies. 
However, the difference between the results of the \citet{Kobulnicky:04a} method 
and that of the \citet{Tremonti:04a} method is small in high-metallicity range, 
and the \citet{Kobulnicky:04a} method tends to result in higher-metallicity 
in low-metallicity range compared to other methods \citep{Kewley:08a}. 
Hence the difference between the metallicity calibration we use 
and that used in \citet{Nagao:06a} would not confuse the separation of the two branches. 

We also estimate star formation rate (SFR) of the host galaxy from the emission line flux. 
Following H$\beta$--SFR relation used in \citet{Savaglio:09a} in which the IMF of \citet{Baldry:03a} is assumed, 
SFR of the host galaxy is 1.88 $M_\odot$/yr, 
or 3.38 $M_\odot$/yr when re-scaled with the Salpeter IMF 
which is consistent to the stellar mass measured in the SED fitting. 
The specific SFR $0.98$ Gyr$^{-1}$ is typical of long GRB host galaxies 
with log~$M_\star/M_\odot \sim 9.5$ \citep{Savaglio:09a}. 

\begin{figure}
  \begin{center}
    \FigureFile(80mm,80mm){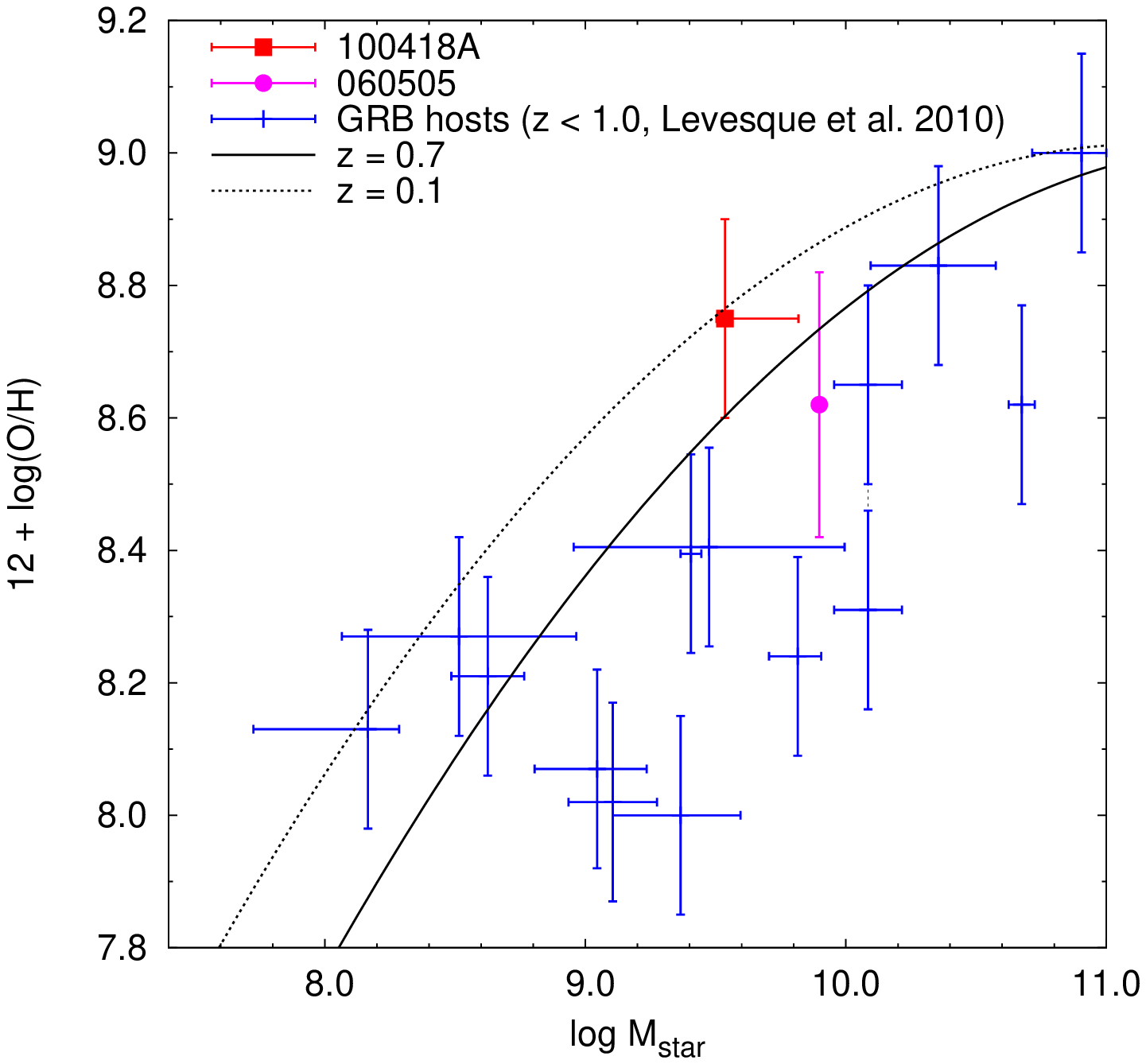}
  \end{center}
  \caption{The stellar mass and the metallicity of the host galaxy of GRB~100418A (red square). 
  Stellar mass and metallicity of host galaxy of GRB~060505 \citep[magenta circle]{Levesque:07a,Thone:08a}, 
  and other GRB host galaxies at $z<1.0$ \citep[blue crosses]{Levesque:10a} are plotted together. 
  Mass--metallicity relation of field galaxies at redshfit 0.1 and 0.7 is shown for comparison \citep{Savaglio:05a}. 
  All metallicities are calibrated with the method of \citet{Kobulnicky:04a}, 
  and stellar masses are re-scaled with the Salpeter IMF. 
  A colored version of the figure is available in the online journal. }\label{fig:hostprop}
\end{figure}

\section{Discussion}

We have presented the results of the search for SN component 
associated with GRB~100418A, with FOCAS on Subaru telescope. 
Both imaging and spectroscopic observation show no evidence of SN. 
Comparison of our $i-$band magnitude to pre-burst data in SDSS catalog 
puts upper limit on afterglow plus SN luminosity, $M_{i,{\rm obs}} > -19.1$. 
Our spectrum shows no SN feature above noise level. 
Assuming that the SN is fainter than the noise level, 
we estimate the upper limit on the SN to be $M_{Ic,{\rm obs}} > -17.2$ (4400--5500~\AA\ in the restframe). 
This limit is about 4 magnitude brighter 
compared to the case of GRB~060505 and GRB~060614 at $z\sim0.1$, 
due to higher redshift of GRB~100418A ($z=0.624$) and brightness of the host galaxy. 
However, it is still comparable to faintest type Ic SNe 
including SN~2002ap which is the faintest broad lined type Ic SN ever observed 
\citep{Drout:11a,Richardson:09a,Mazzali:02a}. 

We have also estimated properties of the host galaxy of GRB~100418A. 
The results of the SED fitting and the emission line diagnostic 
show that the host galaxy has larger stellar mass and higher-metallicity 
than majority of long GRB host galaxies at $z < 1.0$ (figure~\ref{fig:hostprop}). 
It is notable that among three long GRBs with significant limit on their SN component, 
two bursts occurred in galaxies with 12+log(O/H) $> 8.6$ (GRB~060505 and GRB~100418A). 
The other one burst GRB~060614 occurred in a very faint galaxy \citep{Gal-Yam:06a}, 
and metallicity of the host galaxy is not measured. 

\citet{deUgartePostigo:12a} reports possible detection 
of the SN component of GRB~100418A at 28 days after the burst, 
which is faint and agrees with our upper limit. 
The possible SN association suggests that GRB~100418A maybe originate from a collapsar. 
If their possible detection is a real event, 
the SN associated with GRB~100418A is the faintest SN associated with a GRB \citep{Richardson:09a}, 
except another possibly detected SN associated with GRB~101225A 
(\cite{Thone:11a}, but see also \cite{Campana:11a}). 

The specific SFR of the host galaxy of GRB~100418A 
is comparable to other long GRB host galaxies suggesting collapsar origin of this burst, 
while short GRBs occur in galaxies with lower specific SFR than host galaxies of long bursts \citep{Berger:09a}. 
However, it should be noted that a short GRB (XRF) 050416 occurred 
in a galaxy with similar $M_\star$ and SFR to the host galaxy of GRB~100418A \citep{Soderberg:07a}. 

\begin{figure}
  \begin{center}
    \FigureFile(80mm,80mm){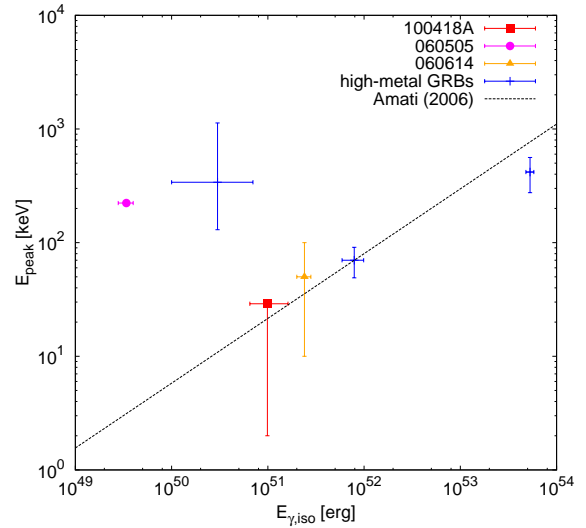}
  \end{center}
  \caption{
  $E_{\rm peak}$ and $E_{\rm \gamma,iso}$ of long GRBs without SNe 
  and long GRBs in high-metallicity host galaxies at $z < 1.0$. 
  GRB~100418A, 060505 and 060614 are shown with filled square (red), 
  circle (magenta), and triangle (orange), respectively. 
  Other bursts are shown with crosses (blue). 
  The $E_{\rm peak}$ and the $E_{\rm \gamma,iso}$ of GRB~100418A is taken from \citet{Marshall:11a}. 
  We take data for the rest of the bursts from \citet{Zhang:09a} and \citet{Amati:07a}. 
  The relation between $E_{\rm peak}$ and $E_{\rm \gamma,iso}$ of long GRBs 
  presented in \citet{Amati:06a} is plotted together. 
  A colored version of the figure is available in the online journal. }\label{fig:EpEi}
\end{figure}

In figure~\ref{fig:EpEi}, we show the spectral peak and the isotropic equivalent gamma-ray energy 
of GRB~100418A \citep{Marshall:11a} together with those of GRB~060505, 060614 
and other GRBs at $z<1.0$ whose host galaxies are metal rich. 
GRB~100418A and 060614 agree with the $E_{\rm peak}$--$E_{\rm \gamma,iso}$ relation of long GRBs by \citet{Amati:06a}, 
while GRB~060505 is in upper-left of the relation where most short bursts reside \citep{Amati:07a}. 
A caveat on this plot is that the $E_{\rm peak}$ of GRB~100418A is calculated in \citet{Marshall:11a} 
using a relation between photon index in {\it Swift} Burst Alert Telescope 
and $E_{\rm peak}$ \citep{Sakamoto:09a}, which is not tested against short GRBs. 
Thus some observational features suggest GRB~100418A originates from a collapsar, 
however none of them are conclusive. 

The sample of long GRBs without bright SNe 
and long GRBs in high-metallicity galaxies is still very small, 
and it is difficult to draw robust conclusions on their nature. 
However, localization of a long GRB without a bright SN in a high-metallicity galaxy 
suggests some interesting possibilities. 
If long GRBs without observable SNe originate from collapsars 
similarly to GRBs with hypernove, many of which occur in low-metallicity galaxies. 
The possible high-metallicity environments of long GRBs without SN features suggest 
environmental effect on explosion scenario in which a collapsar produce no observable SN 
(e.g., a collapsar with a fallback SN, \cite{Woosley:95a,Iwamoto:05a,Moriya:10a}). 
However, it should be noted that metallicity of GRB host galaxies 
maybe systematically different from that of direct environments of GRBs \citep{Niino:11a}.  

\citet{Tominaga:07a} suggested that GRBs associated with hypernovae 
and GRBs without SN features belong to a continuous population, 
predicting the existence of GRBs associated with sub-luminous or faint SNe 
with the $^{56}$Ni mass $\sim 10^{-3}$--$10^{-1}M_\odot$. 
The limit on the SN associated with GRB~100418A based on the noise spectrum 
corresponds to 15\% of the luminosity of SN~1998bw at the peak of its light curve (figure~\ref{fig:SNlimit}). 
Assuming that the mass of $^{56}$Ni in SN ejecta is proportional to the luminosity of the SN, 
the mass of $^{56}$Ni ejected from GRB~100418A is $\lesssim 10^{-1}M_\odot$ 
(the $^{56}$Ni mass ejected from GRB~980425/SN~1998bw is $\sim 0.5M_\odot$, \cite{Iwamoto:98a,Nomoto:01a}). 
In the jet-induced hypernova models by \citet{Tominaga:07a}, 
the ejected amount of $^{56}$Ni is smaller if the jet formation is slower
so that the ram pressure of the jet is lower. 

If the long GRBs without observable SNe occur from non-collapsar origin, 
many of the long GRBs that occur in high-metallicity galaxy may be non-collapsar GRBs. 
In the five GRBs whose host galaxies have spectroscopically measured metallicity 12+log(O/H) $> 8.6$ at $z<1.0$, 
two are without SN component to significant limit, and the other three don't have confirmed SN component. 
Considering the claimed low-metallicity preference of the collapsar model 
(\cite{Yoon:05a, Woosley:06a}; see, however, \cite{Fryer:07a}),  
these observations suggests significant contamination of non-collapsar originate events 
[e.g., GRBs originates in merger of double compact object binary, so called short GRBs or Type I GRBs \citep{Zhang:09a}]
in the sample of $T_{90} > 2$ sec GRBs in high-metallicity galaxies, 
and possibly also those events in low-metallicity galaxies. 

\bigskip
We thank A. de Ugarte Postigo, G. Leloudas and T. Sakamoto for helpful discussions. 
Thanks are also due to our referee whose helpful advice largely improved this paper. 
YN is supported by the Grant-in-Aid for JSPS Fellows.


\begin{thebibliography}{}
\bibitem[Amati(2006)]{Amati:06a} Amati, L.\ 2006, \mnras, 372, 233 
\bibitem[Amati et al.(2007)]{Amati:07a} Amati, L., Della Valle, M., Frontera, F., 
  Malesani, D., Guidorzi, C., Montanari, E., \& Pian, E.\ 2007, \aap, 463, 913 
\bibitem[Antonelli et al.(2010)]{Antonelli:10a} Antonelli, L.~A., et al.\ 2010, GRB Coordinates Network, 10620, 1 
\bibitem[Baldry \& Glazebrook(2003)]{Baldry:03a} Baldry, I.~K., \& Glazebrook, K.\ 2003, \apj, 593, 258 
\bibitem[Berger(2009)]{Berger:09a} Berger, E.\ 2009, \apj, 690, 231 
\bibitem[Bikmaev et al.(2010a)]{Bikmaev:10a} Bikmaev, I., et al.\ 2010, GRB Coordinates Network, 10700, 1 
\bibitem[Bikmaev et al.(2010b)]{Bikmaev:10b} Bikmaev, I., et al.\ 2010, GRB Coordinates Network, 10726, 1 
\bibitem[Bruzual \& Charlot(2003)]{Bruzual:03a} Bruzual, G., \& Charlot, S.\ 2003, \mnras, 344, 1000 
\bibitem[Caito et al.(2009)]{Caito:09a} Caito, L., et al.\ 2009, \aap, 498, 501 
\bibitem[Calzetti et al.(2000)]{Calzetti:00a} Calzetti, D., Armus, L., Bohlin, R.~C., 
  Kinney, A.~L., Koornneef, J., \& Storchi-Bergmann, T.\ 2000, \apj, 533, 682 
\bibitem[Campana et al.(2011)]{Campana:11a} Campana, S., et al.\ 2011, \nat, 480, 69 
\bibitem[Cucchiara \& Fox(2010)]{Cucchiara:10a} Cucchiara, A., \& Fox, D.~B.\ 2010, GRB Coordinates Network, 10624, 1 
\bibitem[Della Valle et al.(2006)]{DellaValle:06a} Della Valle, M., et al.\ 2006, \nat, 444, 1050 
\bibitem[de Ugarte Postigo et al.(in\ prep.)]{deUgartePostigo:12a} de Ugarte Postigo, A., et al.\ in preparation
\bibitem[Drout et al.(2011)]{Drout:11a} Drout, M.~R., et al.\ 2011, \apj, 741, 97 
\bibitem[Filgas et al.(2010)]{Filgas:10a} Filgas, R., Klose, S., \& Greiner, J.\ 2010, GRB Coordinates Network, 10617, 1 
\bibitem[Fryer et al.(2007)]{Fryer:07a} Fryer, C.~L., et al.\ 2007, \pasp, 119, 1211 
\bibitem[Fynbo et al.(2006)]{Fynbo:06a} Fynbo, J.~P.~U., et al.\ 2006, \nat, 444, 1047 
\bibitem[Gal-Yam et al.(2006)]{Gal-Yam:06a} Gal-Yam, A., et al.\ 2006, \nat, 444, 1053 
\bibitem[Galama et al.(1998)]{Galama:98a} Galama, T.~J., et al.\ 1998, \nat, 395, 670 
\bibitem[Gehrels et al.(2006)]{Gehrels:06a} Gehrels, N., et al.\ 2006, \nat, 444, 1044 
\bibitem[Holland et al.(2010)]{Holland:10a} Holland, S.~T., Marshall, F.~E., Page, M., de Pasquale, M., 
  \& Siegel, M.~H.\ 2010, GRB Coordinates Network, 10661, 1 
\bibitem[Iye et al.(2004)]{Iye:04a} Iye, M., et al.\ 2004, \pasj, 56, 381 
\bibitem[Iwamoto et al.(1998)]{Iwamoto:98a} Iwamoto, K., et al.\ 1998, \nat, 395, 672 
\bibitem[Iwamoto et al.(2005)]{Iwamoto:05a} Iwamoto, N., Umeda, H., 
  Tominaga, N., Nomoto, K., \& Maeda, K.\ 2005, Science, 309, 451 
\bibitem[Izotov et al.(2006)]{Izotov:06a} Izotov, Y.~I., Stasi{\'n}ska, G., Meynet, G., Guseva, N.~G., 
  \& Thuan, T.~X.\ 2006, \aap, 448, 955 
\bibitem[Kashikawa et al.(2002)]{Kashikawa:02a} Kashikawa, et al.\ 2002, \pasj, 54, 819 
\bibitem[Kewley \& Ellison(2008)]{Kewley:08a} Kewley, L.~J., \& Ellison, S.~L.\ 2008, \apj, 681, 1183 
\bibitem[Kobulnicky \& Kewley(2004)]{Kobulnicky:04a} Kobulnicky, H.~A., \& Kewley, L.~J.\ 2004, \apj, 617, 240 
\bibitem[Levesque \& Kewley(2007)]{Levesque:07a} Levesque, E.~M., \& Kewley, L.~J.\ 2007, \apjl, 667, L121 
\bibitem[Levesque et al.(2010)]{Levesque:10a} Levesque, E.~M., 
  Kewley, L.~J., Berger, E., \& Zahid, H.~J.\ 2010, \aj, 140, 1557 
\bibitem[Lu et al.(2008)]{Lu:08a} Lu, Y., Huang, Y.~F., \& Zhang, S.~N.\ 2008, \apj, 684, 1330 
\bibitem[Malesani(2010)]{Malesani:10a} Malesani, D.\ 2010, GRB Coordinates Network, 10621, 1 
\bibitem[Marshall et al.(2011)]{Marshall:11a} Marshall, F.~E., et al.\ 2011, \apj, 727, 132 
\bibitem[Marshall et al.(2010a)]{Marshall:10a} Marshall, F.~E., et al.\ 2010, GRB Coordinates Network, 10612, 1 
\bibitem[Marshall \& Holland(2010b)]{Marshall:10b} Marshall, F.~E., 
  \& Holland, S.~T.\ 2010, GRB Coordinates Network, 10720, 1 
\bibitem[Mazzali et al.(2002)]{Mazzali:02a} Mazzali, P.~A., et al.\ 2002, \apjl, 572, L61 
\bibitem[McBreen et al.(2008)]{McBreen:08a} McBreen, S., et al.\ 2008, \apjl, 677, L85 
\bibitem[Moriya et al.(2010)]{Moriya:10a} Moriya, T., Tominaga, N., Tanaka, M., 
  Nomoto, K., Sauer, D.~N., Mazzali, P.~A., Maeda, K., \& Suzuki, T.\ 2010, \apj, 719, 1445 
\bibitem[Nagao et al.(2006)]{Nagao:06a} Nagao, T., Maiolino, R., \& Marconi, A.\ 2006, \aap, 459, 85 
\bibitem[Niino(2011)]{Niino:11a} Niino, Y.\ 2011, \mnras, 417, 567 
\bibitem[Nomoto et al.(2001)]{Nomoto:01a} Nomoto, K., 
  Mazzali, P.~A., Nakamura, T., Iwamoto, K., Danziger, I.~J., \& Patat, F. 
  \ 2001, Supernovae and Gamma-Ray Bursts: the Greatest Explosions since the Big Bang, 144 
\bibitem[Ofek et al.(2007)]{Ofek:07a} Ofek, E.~O., Cenko, S.~B., Gal-Yam, A., Frail, D., 
  Kasliwal, M.~M., Kulkarni, S.~R., \& Waxman, E.\ 2007, \apj, 662, 1129 
\bibitem[Perley et al.(2010)]{Perley:10a} Perley, D.~A., et al.\ 2010, GRB Coordinates Network, 10727, 1 
\bibitem[Richardson(2009)]{Richardson:09a} Richardson, D.\ 2009, \aj, 137, 347 
\bibitem[Rumyantsev \& Pozanenko(2010a)]{Rumyantsev:10a} Rumyantsev, V., 
  \& Pozanenko, A.\ 2010, GRB Coordinates Network, 10783, 1 
\bibitem[Rumyantsev \& Pozanenko(2010b)]{Rumyantsev:10b} Rumyantsev, V., 
  \& Pozanenko, A.\ 2010, GRB Coordinates Network, 10883, 1 
\bibitem[Sakamoto et al.(2009)]{Sakamoto:09a} Sakamoto, T., et al.\ 2009, \apj, 693, 922 
\bibitem[Salpeter(1955)]{Salpeter:55a} Salpeter, E.~E.\ 1955, \apj, 121, 161 
\bibitem[Savaglio et al.(2009)]{Savaglio:09a} Savaglio, S., Glazebrook, K., \& Le Borgne, D.\ 2009, \apj, 691, 182 
\bibitem[Savaglio et al.(2005)]{Savaglio:05a} Savaglio, S., et al.\ 2005, \apj, 635, 260 
\bibitem[Sawicki(2012)]{Sawicki:12a} Sawicki, M. 2012, submitted to \pasp
\bibitem[Schlegel et al.(1998)]{Schlegel:98a} Schlegel, D.~J., Finkbeiner, D.~P., \& Davis, M.\ 1998, \apj, 500, 525 
\bibitem[Smith et al.(2002)]{Smith:02a} Smith, J.~A., et al.\ 2002, \aj, 123, 2121 
\bibitem[Soderberg et al.(2007)]{Soderberg:07a} Soderberg, A.~M., et al.\ 2007, \apj, 661, 982 
\bibitem[Stetson(2000)]{Stetson:00a} Stetson, P.~B.\ 2000, \pasp, 112, 925 
\bibitem[Th{\"o}ne et al.(2011)]{Thone:11a} Th{\"o}ne, C.~C., et al.\ 2011, \nat, 480, 72 
\bibitem[Th{\"o}ne et al.(2008)]{Thone:08a} Th{\"o}ne, C.~C., et al.\ 2008, \apj, 676, 1151 
\bibitem[Tominaga et al.(2007)]{Tominaga:07a} Tominaga, N., Maeda, K., Umeda, H., 
  Nomoto, K., Tanaka, M., Iwamoto, N., Suzuki, T., \& Mazzali, P.~A.\ 2007, \apjl, 657, L77 
\bibitem[Tremonti et al.(2004)]{Tremonti:04a} Tremonti, C.~A., et al.\ 2004, \apj, 613, 898 
\bibitem[Ukwatta et al.(2010)]{Ukwatta:10a} Ukwatta, et al.\ 2010, GRB Coordinates Network, 10615, 1 
\bibitem[Volnova et al.(2010)]{Volnova:10a} Volnova, A., Msu, S., Ibrahimov, M., Karimov, R., 
  \& Pozanenko, A.\ 2010, GRB Coordinates Network, 10821, 1 
\bibitem[Woosley \& Weaver(1995)]{Woosley:95a} Woosley, S.~E., \& Weaver, T.~A.\ 1995, \apjs, 101, 181 
\bibitem[Woosley \& Heger(2006)]{Woosley:06a} Woosley, S.~E., \& Heger, A.\ 2006, \apj, 637, 914 
\bibitem[Xu et al.(2009)]{Xu:09a} Xu, D., et al.\ 2009, \apj, 696, 971 
\bibitem[Yabe et al.(2009)]{Yabe:09a} Yabe, K., Ohta, K., Iwata, I., Sawicki, M., Tamura, N., 
  Akiyama, M., \& Aoki, K.\ 2009, \apj, 693, 507 
\bibitem[Yoon \& Langer(2005)]{Yoon:05a} Yoon, S.-C., \& Langer, N.\ 2005, \aap, 443, 643 
\bibitem[Zhang et al.(2009)]{Zhang:09a} Zhang, B., et al.\ 2009, \apj, 703, 1696 
\end{thebibliography}
\end{document}